# Classified Ads Harvesting Agent and Notification System

Razvi Doomun*, Lollmahamod N., Auleear Nadeem, Mozafar Aukin

*Faculty of Engineering
University of Mauritius, Reduit,
E-mail : r.doomun@uom.ac.mu*

**ABSTRACT**

*The shift from an information society to a knowledge society require rapid information harvesting, reliable search and instantaneous on demand delivery. Information extraction agents are used to explore and collect data available from Web, in order to effectively exploit such data for business purposes, such as automatic news filtering, advertisement or product searching and price comparing. In this paper, we develop a real-time automatic harvesting agent for adverts posted on Servihoo web portal and an SMS-based notification system. It uses the URL of the web portal and the object model, i.e., the fields of interests and a set of rules written using the HTML parsing functions to extract latest adverts information. The extraction engine executes the extraction rules and stores the information in a database to be processed for automatic notification. This intelligent system helps to tremendously save time. It also enables users or potential product buyers to react more quickly to changes and newly posted sales adverts, paving the way to real-time best buy deals.*

**Keywords**

*Information extraction, Web agents, WebL*

## 1.0 INTRODUCTION

Given the enormous growth and great success public information sources on the World Wide Web (WWW), it is increasingly attractive and important to extract data from these sources and make it available for further processing by end users and application programs. The ability to instantaneously access critical Web information anytime, anywhere, and from any device is essential for a variety of tasks, such as information retrieval for business intelligence, event monitoring for stock market, and shopping comparison for e-commerce. Extraction tools using precise navigation and extraction rules greatly reduce the time spent on systematic collection efforts. The most popular applications for information extraction tools remain competitive intelligence gathering and market research, but there are new applications emerging such as deep Web price gathering, primary research, content aggregation for information portals, scientific research and business activity monitoring.

A lot of work has been carried out into the idea of using agents to aid e-commerce, the majority of the attention being focused on B2B agents, with B2C agents receiving a little attention. Sen and Hernandez (2000) discuss the fact that many e-businesses have "seller's agents" whose function it is to push merchandise or services to customers, and there are also "buyer's agents" whose goal is to best serve the user's interests. Maes (1994) discusses how agents used as "personal assistants" that collaborate with the user can be used to reduce work carried out by the user. They can also be used to help with information overload by learning a user's preferences and filtering information presented to the user accordingly. Conceptually, this is similar to our proposed approach. Recommendation agents are agents that calibrate a model of user preferences and use that model to make personalized product recommendations based upon these inferred preferences.

The major objective of this system is to automate the process of extracting unorganised and unstructured information from Servihoo[1] Web portal. We develop a system that periodically searches and extracts the latest products sale, services and adverts information posted on the 'classified ads' web pages of Servihoo web portal. The system automatically processes harvested information according to pre-defined rules and present it back by sending SMS notifications of desired adverts in a timely manner to users who are registered with the system's service. The remainder of this paper is structured as follows: Section 2.0 provides an overview of related work and literature study of information extraction. Section 3.0 describes the object model of information extraction. Section 4.0 is a detail discussion of Web Language. The system architecture is presented in section 5.0. We then discuss and conclude in section 6.0 and 7.0.

## 2.0 RELATED WORK

Information extraction (IE) is a form of shallow document processing task that involves populating a database by automatically obtaining particular fragment of a document that is relevant or of interest. In order to cope with the structural heterogeneity inherent in different information sources, the IE systems rely on a set of extraction rules tailored to a specific information source, often called wrappers, to identify the relevant information to be extracted (Kushmerick, 2000).

---

[1] http://www.servihoo.com

The WysiWyg Web Wrapper Factory (W4F) is a set of tools for automatically generating web wrapper (Sahuguet & Azavant, 2000). It contains a declarative language for extracting data from web pages using extraction rules. W4F does extraction by using an HTML parser to construct a parse tree following a Document Object Model (DOM). The DOM is basically "a platform- and language neutral interface that will allow program and scripts to dynamically access and update the content, structure and style of documents. Taking advantage of this tree structure, some toolkits are designed to produce wrappers that parse these web pages, treating the web pages as a document tree, usually using the DOM as a basis for their extraction rules. In general, HTML tags can help in many tasks involving natural language processing on the web. A certain number of web sites today make use of the hierarchical relations between various HTML elements in crafting out their web pages (Crescenzi, Mecca & Merialdo, 2001). Wrappers built around this structure have several advantages, including ease of use in writing extraction rules by utilizing the HTML tag hierarchy.

A data extraction tool analyzes the tag structure of an HTML document in order to understand how the data is presented in the web page. Laender et al. (Laender, Ribeiro-Neto, Silva &Teixeira, 2002) categorise a number of toolkits based on the methods used for generating wrappers. These methods include specially designed wrapper development languages and algorithms based on HTML-awareness, induction, modelling, ontology and natural language processing. The toolkits are divided into two basic categories based on commercial and non-commercial availability. The stability and reliability of wrappers is highly dependent on the data extraction methods that the toolkit applies. For example, toolkits that only rely on HTML structures to identify relevant data are very vulnerable to the slightest web site changes and frequent repairs to the wrappers may be necessary. Method combinations provide greater robustness, such as the combination of HTML path structures and pattern recognition methods. Some wrapper languages (e.g. HTML Extraction Language in W4F) require the use of absolute HTML paths that point to the data item to be extracted. An absolute path describes the navigation down an HTML tree, starting from the top of the tree (<HTML> tag) and proceeding towards child nodes that contain the data to be extracted.

Stalker (Muslea, Minton, & Knoblock, 2001) expresses hierarchical extraction wrappers as trees in which internal nodes represent lists of records and leaves represent single fields. The system extracts information by descending the tree to successively refine the document segment to be extracted. At each node, extraction boundaries are defined by disjunctions of so-called linear landmark automata, finite-state machines that recognize sequences of tokens, token classes, and wildcards. These automata are intended to consume the prefixes and suffixes of the desired segment, and are learned using an incremental covering algorithm.

## 3.0 OBJECT MODEL AND EXTRACTION

An object model approach is used to extract information from HTML pages. An object model is an abstract image of a user's interest or requirements on the group of web pages. Thus, each web document is transformed into a parse tree corresponding to its HTML hierarchy according to the DOM. The Servihoo classified ads extraction model is based on multiple extraction rules. Extraction has to capture as much structure as it can from the document, a classified advert is composed of various pieces (title, price, date, contacts, etc.); hence, the extraction should be able to capture them altogether. Multiple criteria extraction locates sequential patterns and recursively searches for similar patterns in the web pages (Chang, Hsu & Lui, 2003; Habegger & Quafafou. 2002).

### 3.1 Search methods

There are different ways to extract the information from the web document namely, Element Search, Pattern Search, PCDATA Search and Sequence Search (W3C DOM Technical Committee, 2003). Element Search returns a piece set of all elements that match a specific name. In the context of information extraction, a piece denotes a region on the page while a piece set refers to a collection of pieces belonging to the same page. In Pattern Search, character patterns that match a regular expression are searched in a web page, ignoring the tag objects in the page. This implies that only the pure text stream is searched. For each occurrence of the pattern, a new piece is created. For PCDATA Search, the PCData function returns a piece set of all text segments that are contained in a page or piece. The name PCData is derived from the term "parsed character data", which denotes the text segments on a page, i.e. what is left over when all markup tags are removed from a page. HTML generated on-the-fly by web servers often contains highly stylized markup patterns without hierarchical structure. The markup might be a linear sequence of elements following each other. For example, we might expect an H1 element, followed by a sequence of characters, followed by a BR element. Given a page and a string describing such a sequence (called a sequence pattern), the Sequence search will return a piece set with all the occurrences of the sequence in the page.

### 3.2 Mapping Information

The information extracted from the web document by the evaluation of the extraction rules needs to be mapped in to a certain structure for further processing. This can be achieved in two ways (Gao & Sterling, 1999). Firstly, declaring a Class and mapping the data extracted in to that class. For example if the data retrieved is on books, a class Book can be defined with its attributes and map the data in to the class. Variables can be defined and store the data in them. Secondly, mapping the data into some XML schema and storing the information in some form of XML documents. The information extracted based on the extraction rules is then stored in a Nested String List (NSL) format for future use.

**4.0 WEB LANGUAGE**

WEB Language (WebL) is a web scripting language for processing documents on the World Wide Web (Hannes Marais & Tom Rodeheffer, 1999). It is well suited for retrieving documents from the web, extracting information from the retrieved documents, and manipulating the contents of documents. In contrast to other general purpose programming languages, WebL is designed for rapid prototyping of WEB computations and is well-suited for the automation of tasks on the WWW. Not only does the WebL language have a built-in knowledge of web protocols like HTTP and FTP, but it also knows how to process documents in plain text, HTML and XML format. WebL's emphasis is on high flexibility and high-level abstractions rather than raw computation speed. It is thus better suited as a rapid prototyping tool than a high-volume production tool. It is implemented as a stand-alone application that fetches and processes web pages according to programmed scripts.

In addition to conventional features that would be expected from most languages, the WebL computation model is based on two new concepts, namely **service combinators** and **markup algebra**. A typical WebL program fetches documents from the Web, extracts and refines data from the documents, and then converts the resulting data back into a document on the Web, as shown in figure 1.

WebL (Kistler & Marais, 1998) consists of an embedding of the service combinator algebra. The service combinator algebra is a formalism intended to make the communications aspect of Web computations more reliable, by making possible programs that mimic typical human 'Web reflexes'. Services correspond to Web queries, and encapsulate error detection and handling behaviour. Applying combinators to two or more similar services, which may be unreliable, provides a more reliable 'virtual' service. Interestingly, WebL extends the applicability of service combinators to arbitrary computations, and incorporates an exception handling mechanism. However, there are several problems with this extension. Primarily, these are a result of allowing unconstrained update in the presence of combinators, and interplay between the exception mechanism and combinators whereby failure information is lost. The design decisions in WebL annul the effectiveness of the combinator algebra as a means for formal reasoning, and impede the process of backward error recovery.

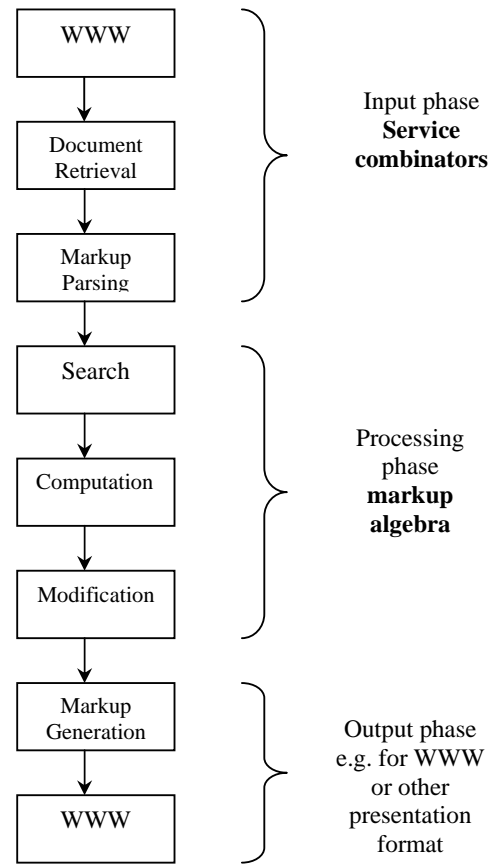

*Figure 1: A model for information extraction application*

Although service combinators cannot make a web-based computation completely failure-proof, it does add a certain amount of robustness to programming on the web. Service combinators are explained in more details in Table 1.

*Table 1: Functions provided by WebL Service Combinators.*

| | |
|---|---|
| **GetURL(url,args)** | Fetch a webpage based on the url |
| **PostURL(url,args)** | Post arguments to a given url |
| **S ? T** | result of S, if S succeeds; otherwise result of T |
| **S \| T** | perform S and T in parallel, result is that of first to succeed; fails if both fail |
| **Timeout (t,S)** | result of S, if S terminates within t milliseconds; otherwise fails |
| **Retry(S)** | result of S, if S succeeds; repeat if S fails |
| **Stall(S)** | never terminates |

**Markup algebra** is a formalism for extracting information from structured text documents and the manipulation of those documents. It consists of functions to extract elements and patterns from web documents, operators to manipulate what has been extracted in this manner, and functions to change a page, for example to insert or delete parts. The functions and operators all work on the high-level concept of a parsed web page. The WebL markup algebra is used for manipulating web pages and extracting data from them. Extracting information may range from simple operations like iterating all the links in a page to more complex operations that fill in Web forms and process the results returned from a server. The markup algebra consists of several operators and functions that operate on pages, tags, pieces and piece sets. There are operators and functions to create or build piece sets from pages or from other piece sets, convert pieces to their string representation, modify the content of a page. After a page is retrieved from the Web and parsed according to its MIME type, the page and its content is accessible for further computation in WebL. The computation that can be performed on a page is determined by the WebL markup algebra.

As the markup algebra of WEBL is based on a set of algebraic operators, a various number of piece set operators and functions are made available to the programmer and they are defined as follows:
- Basic operators (Set Union(P+Q), Set Exclusion (P-Q))
- Positional operators(Indexing P[i], P before/!before Q, P after/!after Q)
- Hierarchical operators (P inside/!inside Q, P contain/! contain Q)
- Regional operators (P without Q, P intersect Q)

## 5.0 SYSTEM ARCHITECTURE

The Information Extraction and SMS Agent consist of three main components: web wrappers (or web agents), server (adverts analysis & SMS agent) and client (web-interface for client users). The system architecture implemented for the information harvesting and notification agent is shown in figure 2. The pseudocode of extraction algorithm is listed in figure 3 and 4. Clients register via a web interface and specify their preferences on which category they want to receive latest information via SMS. Periodically, web agents automatically download latest adverts posted on the web portal. The information analysis and SMS agent analyses the extracted information and compares it with the Preferences database. SMS notification is sent to each user whose preference details match the extracted adverts.

The adverts database consists of tables that will keep information about the different categories found in the "Petites Annonces" section. Information extracted form each category are stored in a specific table. The Preferences database consists of tables that store all the different preferences that exist for each category. A particular car make or model is an example of a car preference. The Clients database keeps all personal details about clients as well as the preferences for which they would register themselves. The SMS database basically consists of two tables, one PendingSMS and the other one SentSMS. Normally before an SMS is sent to a particular client, it is saved in the PendingSMS table and after it has been sent, it is transferred to the SentSMS table.

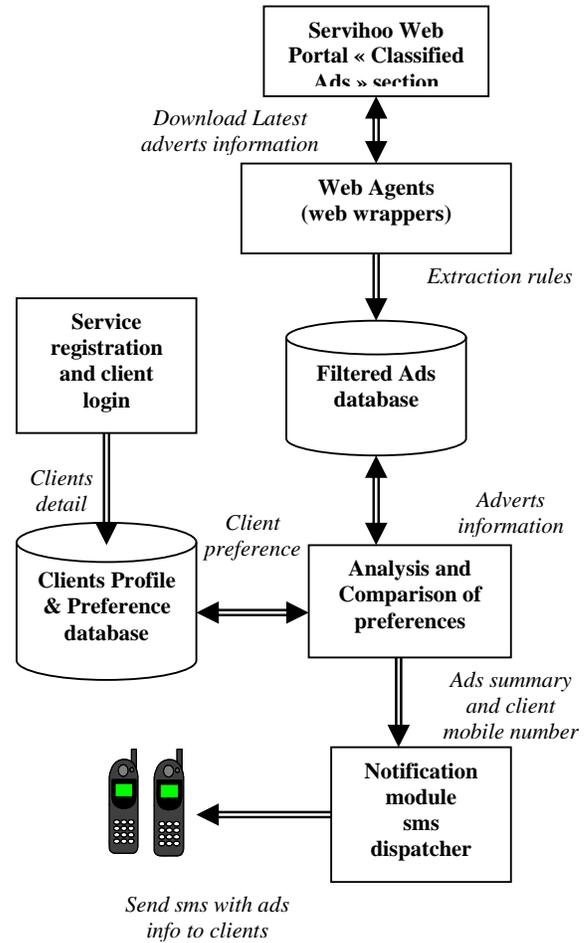

*Figure 2:* Overall system architecture

```
For each link in links-queue
Make connection to link
Pattern Search for current date
Retrieve links that match current date
        If links found then
                Extract Latest Ads Information ()
        Else
                Move next link in links-queue
        End if
End for
```

*Figure 3: Pseudocode for information search*

```
Extract Latest Ads Information ()
Make connection to link and retrieve web page
For each web page retrieved
        Apply Extraction Rules
        Open Database
        Check if information already present
        If information is already present then
                Move next link
        Else
                Save to table in Adverts Database
        End if
        Close database
End for
```

*Figure 4:Pseudocode for extraction*

### 5.1 Description of the components for the Adverts Analysis and SMS Agent

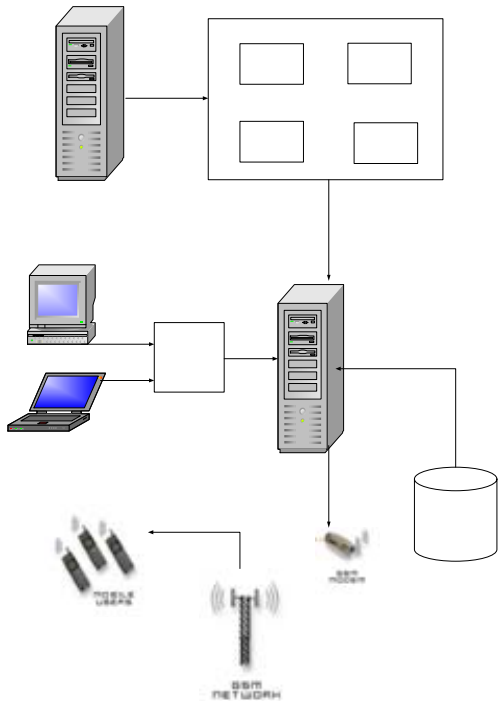

*Figure 5: System components with multiple web agents*

In figure 5, web agents of the different categories extract the latest ads information from the "Petites Annonces" of the Servihoo Web Portal. The latest information of each category are saved to their respective tables in the Adverts databases. After the information extraction process has been completed, the web agents enter a wait state, e.g. 15 minutes.

After the latest ads have been successfully downloaded by the web agents the Adverts Analysis Agent analyses the latest downloaded ads information by first considering each category and then their sub-categories. For each ad's data, of each category the agent looks in the preferences database whether there is a matching preference registered for that ad. For example considering the Car sub-category of the main category Vehicle, a registered preference could be a car with Make- Honda and Model- Civic and if an extracted ad matches this criterion then the agent will retrieve the Preference ID of that ad. After the Preference ID has been successfully retrieved the agent looks in the client database whether we have any client registered for that preference, if found then the agent will check in the SMS database whether we have already sent an SMS on this ad to the client, if not then the agent will save in a table called PendingSMS, the details of the ads, and this process will repeat for each ads downloaded of each category, until all the different categories have been completely analysed

### 5.2 SMS Dispatching Component

Based on the data found on the PendingSMS table in the SMS database, the SMS component retrieves the mobile number of the identified client during the analysis process, compose the message and then dispatch it to the client. After the SMS has been successfully sent, the SMS component will store the message details in a SentSMS table.

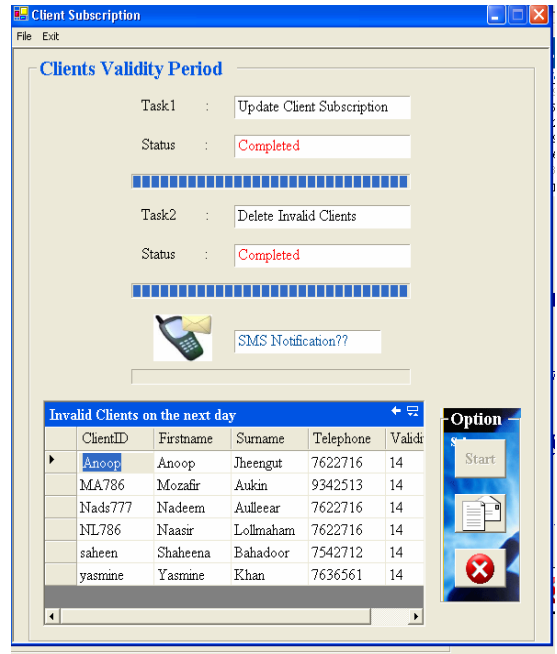

*Figure 6: Application interface with client information*

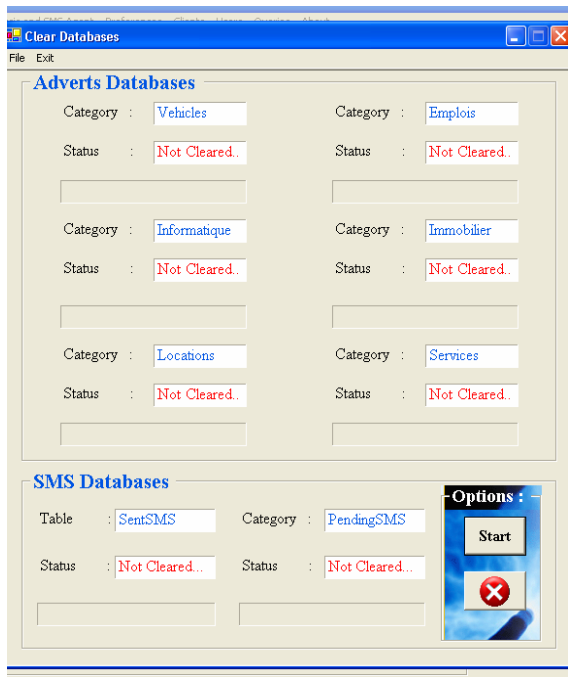

*Figure 7: Interface for starting agent and database status*

For each category identified on the system design, an interface is provided to start each web agent for that category, as shown in figure 7. There are two major methods available for the agent to actually obtain these preferences. The simplest is for the user to communicate their preferences to their agent directly through a simple interface, each time they wish to conduct a preference based search. Alternatively, the agent could learn the user's preferences over time, remembering and inferring them. For simplicity in explaining our new approach, in this paper we assume that the user states their preferences directly to the agent. However, the method of obtaining the user's preferences does not impact on our approach, as long as they are obtained by some means.

## 6.0 DISCUSSION

The system developed is an Intelligent Information Harvester and SMS Agent that is the system once started, automatically launches connection to the Servihoo Web Portal Site, extracts the latest ads information from the "Petites Annonces" section and downloads it to a database. The downloaded information is then dispatched as SMS to registered clients. With such a system, no need for viewers of "Petites Annonces" to each time visit the Servihoo Portal Site and lose time and effort in navigating the classified ads section to obtain latest ads details, what they need to do is just register on the system through the client interface and specify what type of information they want the system to harvest for them and receive the latest ad details on their mobile phone.

Users of the system receive the latest ads details on their mobile phone as the ads have been posted on the classified ads section that is just instantaneously. Clients that have received an SMS on a particular ad, does not receive an SMS of the same ad again when the agent resumes operation. In addition we believe that competition or cooperation of the specialized agents can bring better results than a single process, single representation system.

## 7.0 CONCLUSION

In our approach, the DOM tree is used to perform web adverts extraction by first parsing web pages into DOM trees. Extraction patterns are then specified as paths from the root of the DOM tree to the node containing the text to extract. The technique that we have employed is simple, but effective. Users are automatically notified of desired ads via SMS and they can profit from best deals with fast response. As the structure of a website keeps changing and updating, in the system, links are retrieved dynamically and information is extracted based on Pattern searching methodology and not depending totally on the structure of the HTML document.

In this paper we have designed and implemented a specific area of application (namely adverts notification for e-commerce). The main characteristic of our system is that an agent acts on behalf of the user, matching the user's preferences to the latest harvested adverts in the database, thus ensuring the notification results received are the "best deals" results for the user.

Various directions for future work exist. It might be helpful to develop additional functionalities for application of more complex domains and coorperative multi-agent processing with learning mechanisms.